\documentclass{aastex}          
\usepackage{spr-astr-addons}    
\usepackage{longtable}
\usepackage{lscape}

\begin{document}
%
\title{A study of Intrinsic $\gamma$-ray Emission for  {\it Fermi}/LAT-detected BL Lacs}

\shorttitle{Intrinsic $\gamma$-ray Emission}
\shortauthors{Ye et al.}

\author{X. H. Ye\altaffilmark{1,2,3}} \and \author{X. T. Zeng\altaffilmark{1,2,3}} \and \author{W. X. Yang\altaffilmark{1,2,3}}\and \author{H. S. Huang\altaffilmark{4}}\and \author{Y. H. Xuan\altaffilmark{4}}\and \author{J. W. Huang\altaffilmark{4}}\and \author{Z. Zhang\altaffilmark{1,2,3}}\and \author{Z. Y. Pei\altaffilmark{1,2,3}}\and \author{J. H. Yang\altaffilmark{5,1}}\and \author{J. H. Fan\altaffilmark{1,2,3,*}}
\altaffiltext{*}{Corresponding author: J. H. Fan, fjh@gzhu.edu.cn}
\altaffiltext{1}{Center for Astrophysics, Guangzhou University, Guangzhou 510006, China}
\altaffiltext{2}{Astronomy Science and Technology Research Laboratory of Department of Education of Guangdong Province, Guangzhou 510006, China}
\altaffiltext{3}{Key Laboratory for Astronomical Observation and Technology of Guangzhou, Guangzhou 510006, China}
\altaffiltext{4}{School of Physics and Material Science, Guangzhou University, Guangzhou 510006, China}
\altaffiltext{5}{Department of Physics and Electronics Science, Hunan University of Arts and Science, Changde 415000, China}

\begin{abstract}
BL Lacs are one subclass of  blazars with highly energetic $\gamma$-ray emission, which is strongly boosted by a relativistic beaming effect. The latest catalogue of the 10 years of {\it Fermi}/LAT data \citep{abd2020} and the $\gamma$-ray Doppler factors in \citet{pei2020}  provide us with a large BL Lac sample to study their jet emission morphologies and intrinsic properties. In this paper, we collected a sample of 294 {\it Fermi} BL Lacs and probed the correlations between the $\gamma$-ray emissions and luminosity distances. Our analyses give following conclusions: (1) the observed $\gamma$-ray emissions are really boosted by the $\gamma$-ray Doppler factor, and the intrinsic $\gamma$-ray emissions are closely correlated with luminosity distances.  (2) 
the morphology of jet emissions for HBLs may be continuous,  while that for IBLs may be the case of a moving sphere in the $\gamma$-ray bands.
\end{abstract}


\section{Introduction}\label{s:1}

Active galactic nuclei (AGNs) are  one kind of special galaxies in the Universe. Their bolometric luminosity is much higher than that of normal galaxies. The physical structure of AGNs is that a supermass black hole (SMBH) locates in the centre, surrounded by  an accretion disc with two relativistic jets. The clouds of gas near the SMBH are radiating strong optical and ultraviolent emissions, the so-called broad line region (BLR), while those clouds of gas far away from the SMBH producing narrow lines is called the narrow line region (NLR) \citep{urry1995}. Blazars are a subclass of AGNs with  following observational properties, such as rapid variability superimposed on the slow longterm variation, high and variable polarization, apparent superluminal motion, radio core dominance, and strong $\gamma$-ray emissions \citep{sil1988, wil1992, don1995, fan1998, har1999, xiao2019, abd2020, ote2020, pei2020,  fan2021, rai2021, yuan2021}. Based on the feature of emission lines, blazars can be classified as BL Lacs and flat spectrum radio quasars (FSRQs). The optical spectrum of BL Lacs shows weak or absent emission line feature (equivalent width, EW $< 5 \mathring{A}$), while that of FSRQs has strong emission line feature (EW $\geq 5\mathring{A}$)  \citep{1991ApJ...374..431S, urry1995}.

The spectral energy distributions (SEDs) of blazars show a bimodal structure: the first bump (or lower frequency bump) is from far-infrared to soft X-ray band, which is produced by the synchrotron radiation of electrons within the jet, and the second bump  (or higher frequency bump) from MeV to TeV band  is derived by the inverse Compton (IC) radiation \citep{pad1996, ghi1998, fan2016}. According to the different synchrotron peak frequency of the first bump, BL Lacs can be classified into different subclasses. \citet{pad1996} divided BL Lacs into higher  frequency [log $(\nu$/Hz) $\geq$ 15] peak BL Lacertae objects (HBLs) and lower  frequency [log $(\nu$/Hz) $<$ 15] peak BL Lacertae objects (LBLs). \citet{abd2010} calculated the SEDs using the quasi-simultaneous multiwavelength observations  for a sample of 48 Fermi blazars and proposed to classify the Fermi blazars as low, intermediate and high synchrotron peaked blazars (LSP, ISP, and HSP). In this sense, we have log $(\nu$/Hz) $\geq 15$ for HBLs, 14 $\leq$ log $(\nu$/Hz) $\leq 15$ for intermediate synchrotron peaked BL Lacs (IBLs), log $(\nu$/Hz) $\leq14$ for LBLs. Later on, SEDs were calculated for a large sample of 1425 Fermi blazars in \citet{fan2016}, who also proposed the classification of blazars, namely: log $(\nu$/Hz) $\geq 15.3$ for HBLs, 14 $\leq$ log $(\nu$/Hz) $\leq 15.3$ for IBLs, log $(\nu$/Hz) $\leq 14$, for LBLs.  In this work, we adopted the classification in \citet{fan2016}.

In a beaming model, the observed flux density ($f^{\rm{ob}}$) is enhanced by the extremely relativistic effect or Doppler factor ($\delta$), $f^{\rm{ob}}=\delta^q f^{\rm{in}}$,  in which $f^{\rm{in}}$ is the intrinsic flux density, $q=3+\alpha$ is for the case of a moving sphere, $q=2+\alpha$ for the case of a continuous jet. \citep{lin1985,ghi1993},  here $\alpha$ is a spectral index ($f_{\nu}\propto \nu^{-\alpha} $). Doppler factor ($\delta$) can be determined by an viewing angle ($\theta$) and a bulk velocity in units of speed of light ($\beta=v/c$): $\delta=[\Gamma(1-\beta \cos \theta)]^{-1}$, where the Lorentz factor ($\Gamma$) is also determined by the bulk velocity ($\beta$): $\Gamma=1 / \sqrt{1-\beta^{2}}$. Since it is difficult to obtain the  viewing angle and the bulk velocity, one can not obtain the Doppler factor from observations directly. Fortunately, many indirect methods have been proposed to estimate the  Doppler factor, such as Doppler factor estimation from radio flux density variation or variability brightness temperature  \citep{lah1999, fan2009, hov2009, lio2018},  from the combination between X-ray emissions and synchrotron Self-Compton (SSC) modal \citep{ghi1993}, from pair-production opacity in the high energy emission \citep{mat1993,fan2013, fan2014,pei2020},  from broadband SED fitting \citep{chen2018}. Recently, \citet{zhang2020} proposed a new method using both the $\gamma$-ray luminosity and BLR luminosity to estimate Doppler factor for a sample of 350 blazars. Meanwhile, based on the unification of BL Lacs and their parent population of Fanaroff \& Riley type I (FRI)  radio galaxies and Fanaroff \& Riley type  II with radio morphology (FRII-G) \citep{fan1974}, \citet{ye2021} assumed that  the intrinsic correlation between core luminosity and extended luminosity for BL Lacs is the same as that of their parent radio galaxies (FRI and FRII-G), then estimated the Doppler factor for a sample of 297 BL Lacs.  Due to the different morphologies of jet emissions, the relativistic beaming effect causes the intrinsic $\gamma$-ray flux density to appear amplified in different type (a continuous jet or a case of a moving sphere).  In this paper, our intent is to probe the relations between the $\gamma$-ray flux densities and luminosity distances  for {\it Fermi} BL Lacs to study their morphologies of  jet emissions and intrinsic properties. The samples and results are shown in Sect. 2, and some discussions and conclusions are presented in Sect. 3 and Sect. 4.

\section{Samples and Results}\label{s:2}

\subsection{Samples}\label{s:2.1}
From \citet{pei2020}, we obtained 467 {\it Fermi} BL Lacs.  Then we checked in the latest Fermi/LAT incremental Data Release (4FGL-DR2, for Data Release 2) for the power law spectrum \citep{abd2020} and the NED (NASA/IPAC Extragalactic Database) for available redshift. Finally, we got a sample of 294 {\it Fermi} BL Lacs with Doppler factor and redshift. Out of the 294 BL Lacs, 27 are  LBLs, 135 IBLs, and 132 HBLs. The relevant data are listed in Table \ref{tab1}, in which Col. 1-7 are basic informations and Doppler factors from \citet{pei2020}, Col. 8-13 for the corresponding $\gamma$-ray emissions.

\setlength{\tabcolsep}{1pt}
\begin{table*}
\caption{The properties of whole samples.}
\begin{tabular}{@{}lcccccccccccc@{}}
\hline
\hline
Source  Name&class&$z$ & $\alpha_{\rm{ph}}$  & $N$&$\delta_{\rm{P20}}$&$\delta_{\rm{P20}}$&$f_{\gamma}^{\rm{ob}}$&$f_{\gamma}^{\rm{in2}}$&$f_{\gamma}^{\rm{in3}}$&log L$_{\gamma}^{\rm{ob}}$&log L$_{\gamma}^{\rm{in2}}$&log L$_{\gamma}^{\rm{in3}}$ \\
&&&& &($q$=2) &($q$=3) &($10^{-5}$ pJy) &($10^{-5}$ pJy)&($10^{-5}$ pJy)&erg s$^{-1}$&erg s$^{-1}$&erg s$^{-1}$\\
(1)&(2)&\,(3)&(4)&\,\,(5)&(6)&(7)&(8)&(9)&(10)&(11)&(12)&(13)\\
\hline
4FGL	J0006.3-0620	&	HBL	&	0.347	&	2.17	&	1.26	&	2.95	&	3.61	&	106.50	&	1.82	&	1.17	&	46.74	&	44.97	&	44.78	\\
4FGL	J0008.4-2339	&	IBL	&	0.147	&	1.68	&	1.95	&	2.34	&	2.84	&	610.41	&	37.06	&	26.71	&	46.66	&	45.44	&	45.30	\\
4FGL	J0013.9-1854	&	IBL	&	0.095	&	1.97	&	2.79	&	2.14	&	2.50	&	411.03	&	27.28	&	20.15	&	46.08	&	44.90	&	44.77	\\
4FGL	J0014.2+0854	&	HBL	&	0.163	&	2.50	&	1.82	&	1.60	&	1.73	&	55.30	&	8.06	&	6.62	&	45.71	&	44.88	&	44.79	\\
4FGL	J0018.4+2946	&	HBL	&	0.1	&	1.72	&	2.62	&	2.07	&	2.44	&	763.00	&	67.43	&	50.85	&	46.39	&	45.34	&	45.22	\\
4FGL	J0032.4-2849	&	IBL	&	0.324	&	2.30	&	1.94	&	2.00	&	2.26	&	113.61	&	7.75	&	5.84	&	46.70	&	45.53	&	45.41	\\
4FGL	J0035.2+1514	&	IBL	&	0.25	&	1.84	&	11.04	&	2.97	&	3.74	&	2255.43	&	52.95	&	34.52	&	47.74	&	46.11	&	45.92	\\
4FGL	J0045.7+1217	&	HBL	&	0.228	&	2.00	&	9.46	&	2.56	&	3.09	&	1265.89	&	42.89	&	29.45	&	47.40	&	45.93	&	45.76	\\
4FGL	J0047.9+3947	&	IBL	&	0.252	&	1.98	&	7.40	&	1.86	&	2.11	&	1039.59	&	111.61	&	87.04	&	47.41	&	46.44	&	46.33	\\
4FGL	J0056.3-0935	&	IBL	&	0.103	&	1.87	&	6.78	&	2.14	&	2.51	&	1302.23	&	92.21	&	68.28	&	46.65	&	45.50	&	45.37	\\
4FGL	J0059.3-0152	&	IBL	&	0.144	&	1.79	&	2.05	&	1.97	&	2.28	&	492.12	&	49.47	&	37.98	&	46.54	&	45.55	&	45.43	\\

$\cdots$ & $\cdots$ & $\cdots$ & $\cdots$ & $\cdots$ & $\cdots$ & $\cdots$ & $\cdots$ & $\cdots$ & $\cdots$ &$\cdots$&$\cdots$&$\cdots$\\
\hline
\end{tabular}\label{tab1}

\medskip

\small
Note: Col. 1 is the 4FGL name, Col. 2 classification, LBL for LSP BL Lac, IBL for ISP BL Lac, HBL for HSP BL Lac,  Col. 3 redshift, $z$, Col. 4 photon spectral index, $\alpha_{\rm{ph}}$, Col. 5 the integral photon flux ($N$) for 1 - 100 GeV in units of $10^{-10}$ photons/cm$^2$/s,  Col. 6 the $\gamma$-ray Doppler factor for $q$ = 2 from \citet{pei2020}. Col. 7 the $\gamma$-ray Doppler factors for $q$ = 3 from \citet{pei2020}. Col. 8 the observed flux density, $f_{\gamma}^{\rm{ob}}$ at E = 50 GeV in units of $10^{-5}$ pJy, Col. 9 the intrinsic flux density ($f_{\gamma}^{\rm{in2}}$) for $q$ = 2 in units of $10^{-5}$ pJy, Col. 10 the intrinsic flux density ($f_{\gamma}^{\rm{in3}}$) for $q$ = 3 in units of $10^{-5}$ pJy, Col. 11 the logarithm of observed luminosity in units of erg s$^{-1}$,  Col. 12 the logarithm of intrinsic luminosity for  $q$ = 2 in units of erg s$^{-1}$,  Col. 13 logarithm of the intrinsic luminosity for  $q$ = 3 in units of erg s$^{-1}$.

(A portion is shown here for guidance regarding its form and content. Readers can also contact author to obtain the whole sample.)
\end{table*}

\subsection{Results}\label{s:2.2}
\begin{figure}[bht]
\includegraphics[width=1\columnwidth]{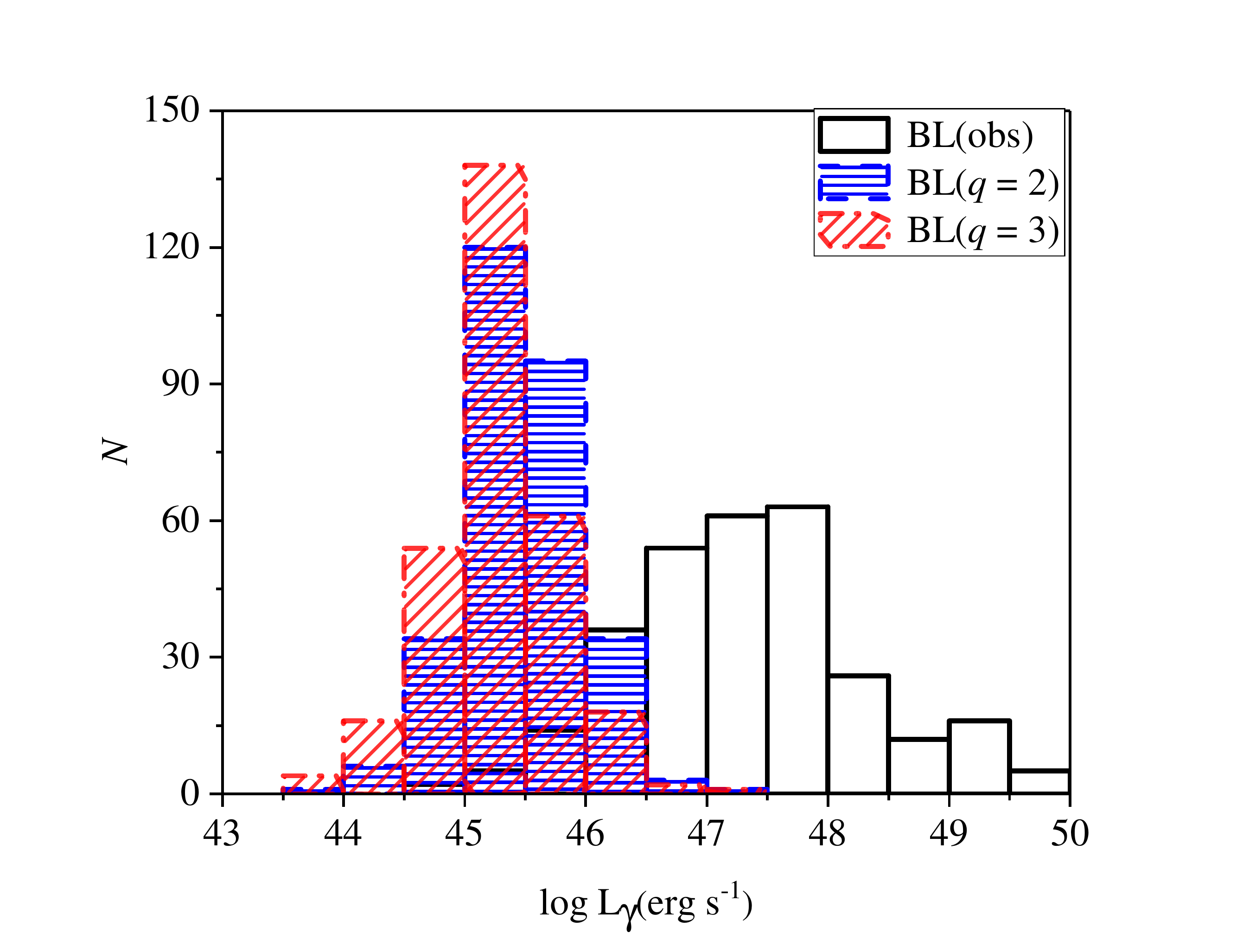}
\caption{Plot of the distributions of  the logarithm of observed and intrinsic luminosities for BL Lacs. The unhatched areas show the observed luminosities. The Areas with horizontal lines show the intrinsic luminosities for $q$ = 2. The Areas with slashes show  the intrinsic luminosities for $q$ = 3.}
\label{fig:1}
\end{figure}

For a $\gamma$-ray BL Lac, when the photon flux is adopted for a power law function, $\mathrm{d} N / \mathrm{d} E=N_{0} E^{-\alpha_{\mathrm{ph}}}$, where $N$ is the integral photon flux in the units of  photons/cm$^2$/s, and $\alpha_{\mathrm{ph}}$ is the photon spectral index, $N_{0}$ can be expressed as \citep{fan2013, xiao2015, lin2017, yang2017}
\begin{equation}
N_{0}=N\cdot (1-\alpha_{\mathrm{ph}})/(E_{\mathrm{U}}^{1-\alpha_{\mathrm{ph}}}-E_{\mathrm{L}}^{1-\alpha_{\mathrm{ph}}})
\end{equation}
then the $\gamma$-ray flux density at energy $E$ in units of pJy ($10^{-12}$ Jy) is
\begin{equation}
f_{\mathrm{E}}=6.63 \times 10^{8} N \cdot \frac{1-\alpha_{\mathrm{ph}}}{E_{\mathrm{U}}^{1-\alpha_{\mathrm{ph}}}-E_{\mathrm{L}}^{1-\alpha_{\mathrm{ph}}}} \cdot E^{1-\alpha_{\mathrm{ph}}}(\mathrm{pJy})    \label{flux}
\end{equation}
$E_{\mathrm{U}}$ = 100 GeV and $E_{\mathrm{L}}$ = 1 GeV in this paper.  The observed $\gamma$-ray flux density is K-corrected by using $f_{\rm{cor}}=f_{\rm{E}}(1+z)^{\alpha_{\gamma}-1}$, here $\alpha_\gamma =\alpha_{\mathrm{ph}}-1$ ($f_{\nu}\propto \nu^{-\alpha_{\gamma}} $) is a $\gamma$-ray spectral index. According to equation (\ref{flux}), we computed the observed flux densities $f_{\rm{E}}$ at E = 50 GeV for the 294 BL Lacs and listed the observed $\gamma$-ray flux densities after K-correction in Col. 8 of Table \ref{tab1}. As well known, the observed $\gamma$-ray flux densities ($f^{\rm{ob}}_\gamma$) are enhanced by a strong beaming effect, expressed as the $\gamma$-ray Doppler factor ($\delta_\gamma$). When the $\gamma$-ray Doppler factors by \citet{pei2020} [hereafter $\delta_{\rm{P20}}$] are adopted to study the intrinsic $\gamma$-ray flux densities, one can get 
\begin{equation}
logf^{\rm{in}}_\gamma=logf_\gamma^{\rm{ob}}-q \log \delta_{\rm{P20}}\label{in_flux}
\end{equation} 
$q$ = 2 + $\alpha_\gamma$ or 3 + $\alpha_\gamma$ (see above). The corresponding intrinsic $\gamma$-ray flux densities for the 294 BL Lacs are  listed in the Col. (9) and (10) of Table \ref{tab1}.  The $\gamma$-ray luminosity can be obtained by a formula, $L_{\gamma}=4\pi d_{\rm{L}}^2\nu_{\gamma}f_{\gamma}$, here $d_{\rm{L}}$ is the luminosity distance determined by: 
\begin{equation}
d_{\mathrm{L}}=(1+z)\frac{ c}{H_{0}} \int_{1}^{1+z} \frac{1}{\sqrt{\Omega_{\mathrm{M}} x^{3}+1-\Omega_ {\mathrm{M}}}} dx \label{dis}
\end{equation}
with $\Omega_{\Lambda}\sim0.692$, $\Omega_{\rm{M}}\sim0.308$ and $H_0=67.8$km s$^{-1}$ Mpc$^{-1}$ \citep{pla2016}. 
The corresponding observed and intrinsic luminosities for the 294 BL Lacs are listed in the Col. (11)-(13) of Table. \ref{tab1} and the corresponding histograms are plotted in Fig. \ref{fig:1}.  
%

The logarithm of observed luminosity in units of erg s$^{-1}$, log (L$_\gamma^{\rm{ob}}$/erg s$^{-1})$, is in a range of log (L$_\gamma^{\rm{ob}}$/erg s$^{-1}$) = 44.59$--$49.70, with a Gaussian average of 47.25 $\pm$ 0.84, and that of intrinsic luminosity for $q$ = 2 is in a range of  log (L$_\gamma^{\rm{in2}}$/erg s$^{-1}$) = 44.00$--$47.08 with a Gaussian average of 45.25 $\pm$ 0.47, while  that of intrinsic luminosity for $q$ = 3 is in a range of  log (L$_\gamma^{\rm{in3}}$/erg s$^{-1}$) = 43.67$--$47.06 with a Gaussian average of 45.25 $\pm$ 0.41. As we can see in Fig. \ref{fig:1}, the distribution of intrinsic luminosities for BL Lacs shows a two dex narrower range than the observed luminosity does. It can be seen that the intrinsic luminosity distributes mainly between 45 and 46 (for q = 2 and 3 case) while the observed luminosity distributes mainly in a larger range. If one assumes that the intrinsic luminosity is constant for BL Lacs, one can expect the logarithm of flux density to follow a linear correlation with the logarithm of luminosity distance, $\log f_{\gamma}= -2.0\log d_{\rm{L}}  + $ const. Based on the assumption, we investigated the relations between the $\gamma$-ray flux densities and luminosity distances for the compiled BL Lacs to study their morphologies of jet emissions and intrinsic properties.

{\it BL Lac:} When a linear relation is adopted to the observed $\gamma$-ray flux densities (in units of pJy) and luminosity distances (in units of Mpc) for 294 BL Lacs, we have 
\begin{equation*}
\log f^{\rm{ob}}_{\rm{BL}}=-(0.16\pm0.05)\log d_{\rm{L}} - (1.77\pm0.17)
\end{equation*}
with a correlation coefficient $r$= -0.174 and a chance probability $p = 2.7\times 10^{-3}$. The slope (-0.16) for the 294 BL Lacs is significantly different with the expected slope of -2.  When the relations for intrinsic flux densities against luminosity distances are considered,  we have
 \begin{equation*}
\log  f^{\rm{in2}}_{\rm{BL}}=-(1.96\pm0.06)\log d_{\rm{L}} + (2.18\pm0.19)
\end{equation*}
with $r$ = -0.890 and $p < 10^{-4}$ for the case of $q=2+\alpha_\gamma$ (the case of a continuous jet), and 
 \begin{equation*}
\log f^{\rm{in3}}_{\rm{BL}}=-(2.16\pm0.06)\log d_{\rm{L}} +(2.61\pm0.20)
\end{equation*}
with $r$  = -0.897 and $p < 10^{-4}$ for the case of $q=3+\alpha_\gamma$ (the case of a moving sphere).

\begin{figure}[bht]
\includegraphics[width=1\columnwidth]{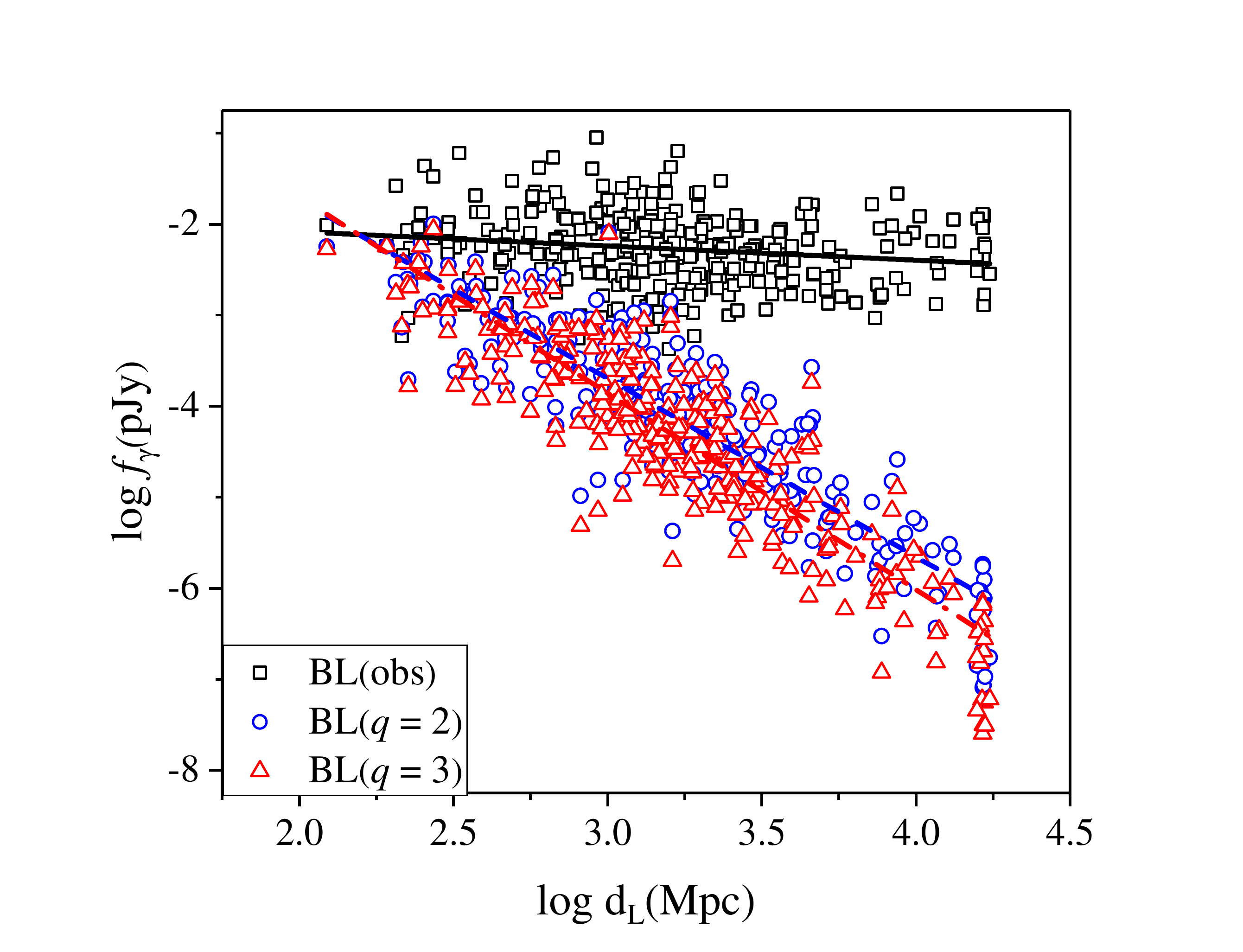}
\caption{Plot of the relations between $\gamma$-ray flux densities and luminosity distances for 294 BL Lac based on the $\gamma$-ray Doppler factor of \citet{pei2020}.  The {\it square} stands for observed $\gamma$-ray flux density sample and the {\it solid line} stands for the best observed fitting regression line, the {\it circle} for continuous jet ($q$ = 2)  and the {\it broken line} for the best ($q$ = 2) fitting regression line, the {\it triangle} for the case of a moving sphere} ($q$ = 3), and the {\it dot-dashed line}  for the best ($q$ = 3) fitting regression line.
\label{fig:2}
\end{figure}

The corresponding results are shown in Fig. \ref{fig:2}.
%

\begin{figure}[bht]
\includegraphics[width=1\columnwidth]{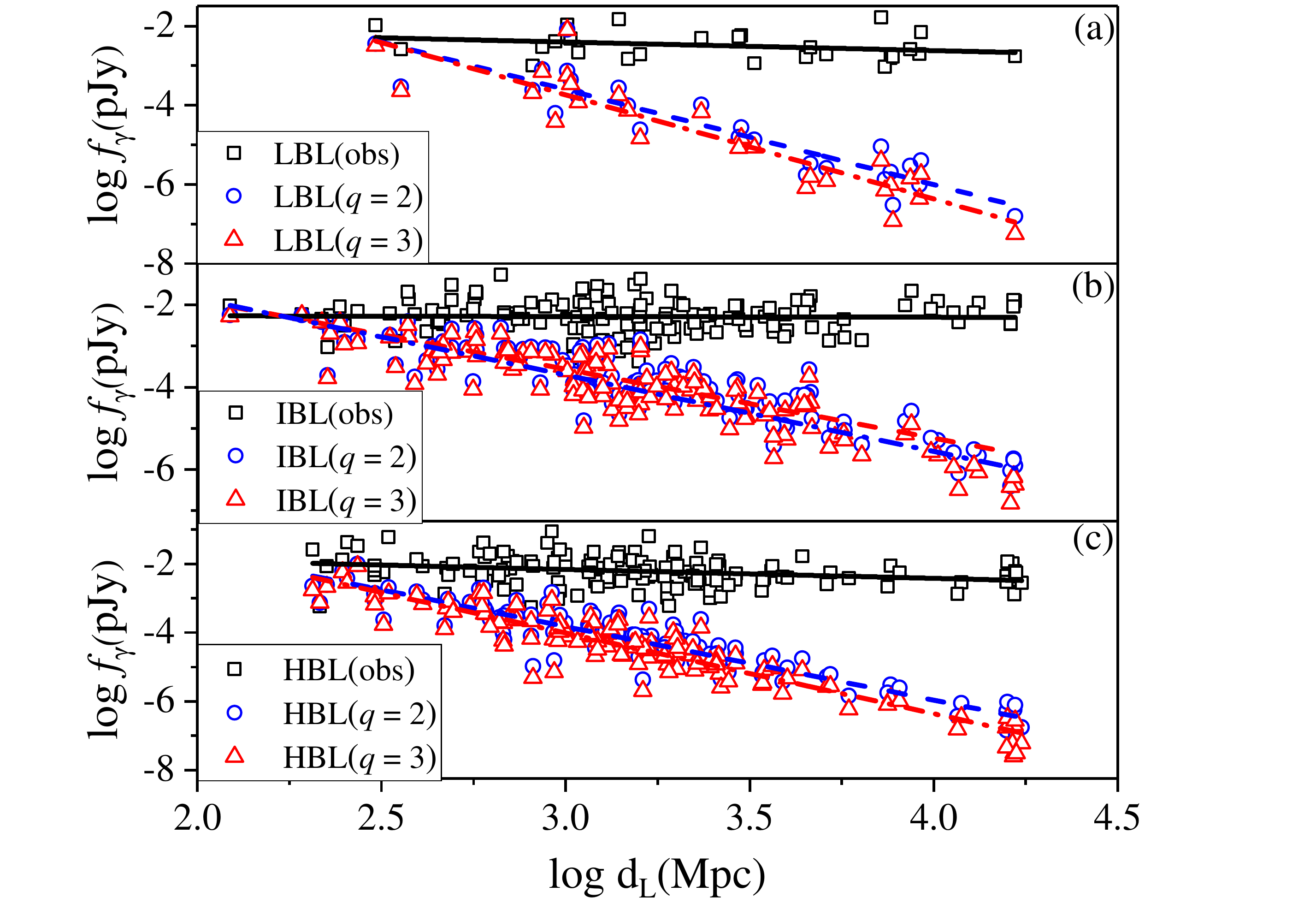}
\caption{Plot of the relations between $\gamma$-ray flux densities and luminosity distances for 27 LBLs, 135 IBLs, and 132 HBLs based on the $\gamma$-ray Doppler factor of \citet{pei2020}. (a): LBLs, (b): IBLs, (c): HBLs. All the representations of labels are the same as Fig. \ref{fig:2}.}
\label{fig:3}
\end{figure}

When we consider the subclasses separately, the linear correlations for LBLs, IBLs, and HBLs are presented in Fig. \ref{fig:3} and Table \ref{tab2}, in which Col. 2-3 are the linear fitting coefficients, Col. 5-6 for  the correlation coefficients and their corresponding chance probabilities. Col. 7-8 for the partial correlation coefficients and their corresponding chance probabilities after removing Doppler factor effect.

\section{Discussions}\label{s:3}
\begin{figure}[h]
\includegraphics[width=1\columnwidth]{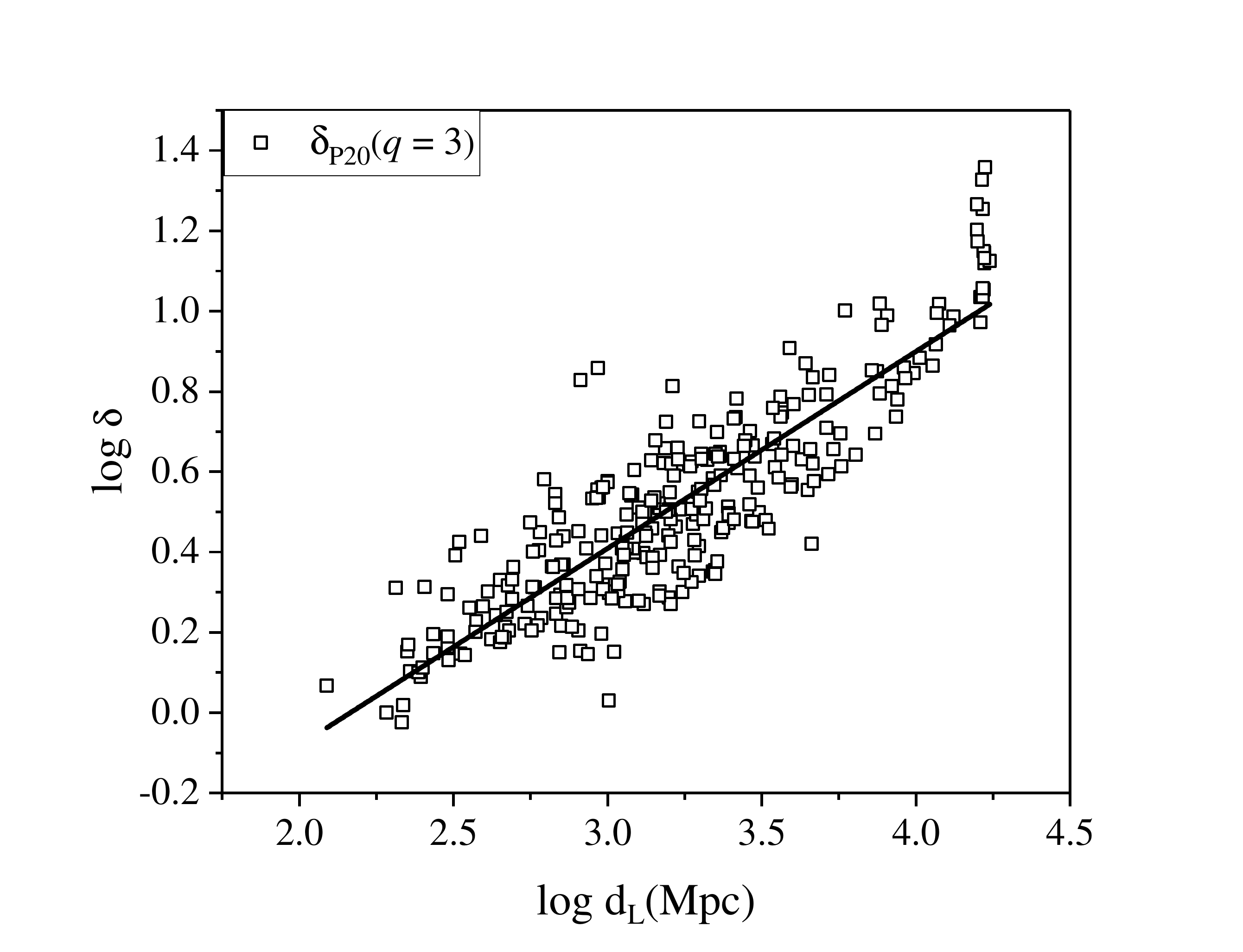}
\caption{Plot of the relation between $\gamma$-ray Doppler factors ($q$ = 3) of \citet{pei2020} and luminosity distances for 294 BL Lacs.}
\label{fig:4}
\end{figure}

Following the works \citep{mat1993, fan2013}, \citet{pei2020} assumed that both the X-ray and the $\gamma$-ray are from the same region,  adopted the variability timescales of one day, and estimated the $\gamma$-ray Doppler factor  for a sample of 809 {\it Fermi}/LAT blazars. The tight correlations of the $\gamma$-ray luminosity against $\gamma$-ray Doppler factor were presented in the Fig. 4 of \citet{pei2020}, who stated that the $\gamma$-ray emissions are strongly beamed and the jet emission morphology may be continuous in $\gamma$-ray band.  From the sample of  \citet{pei2020},  we collected 294 $\gamma$-ray Doppler factors and found that  the $\delta_{\rm{P20}}$ ($q$ = 3) are closely correlated with the  luminosity distances [log $\delta_{\rm{P20}} =(0.49 \pm 0.02)$ log $d_{\rm{L}}-(1.06 + 0.05)$]  with $r$ = 0.875 and $p$ $<$ 10$^{-4}$ as shown in Fig. \ref{fig:4}.

%

\begin{figure}[bht]
\includegraphics[width=1\columnwidth]{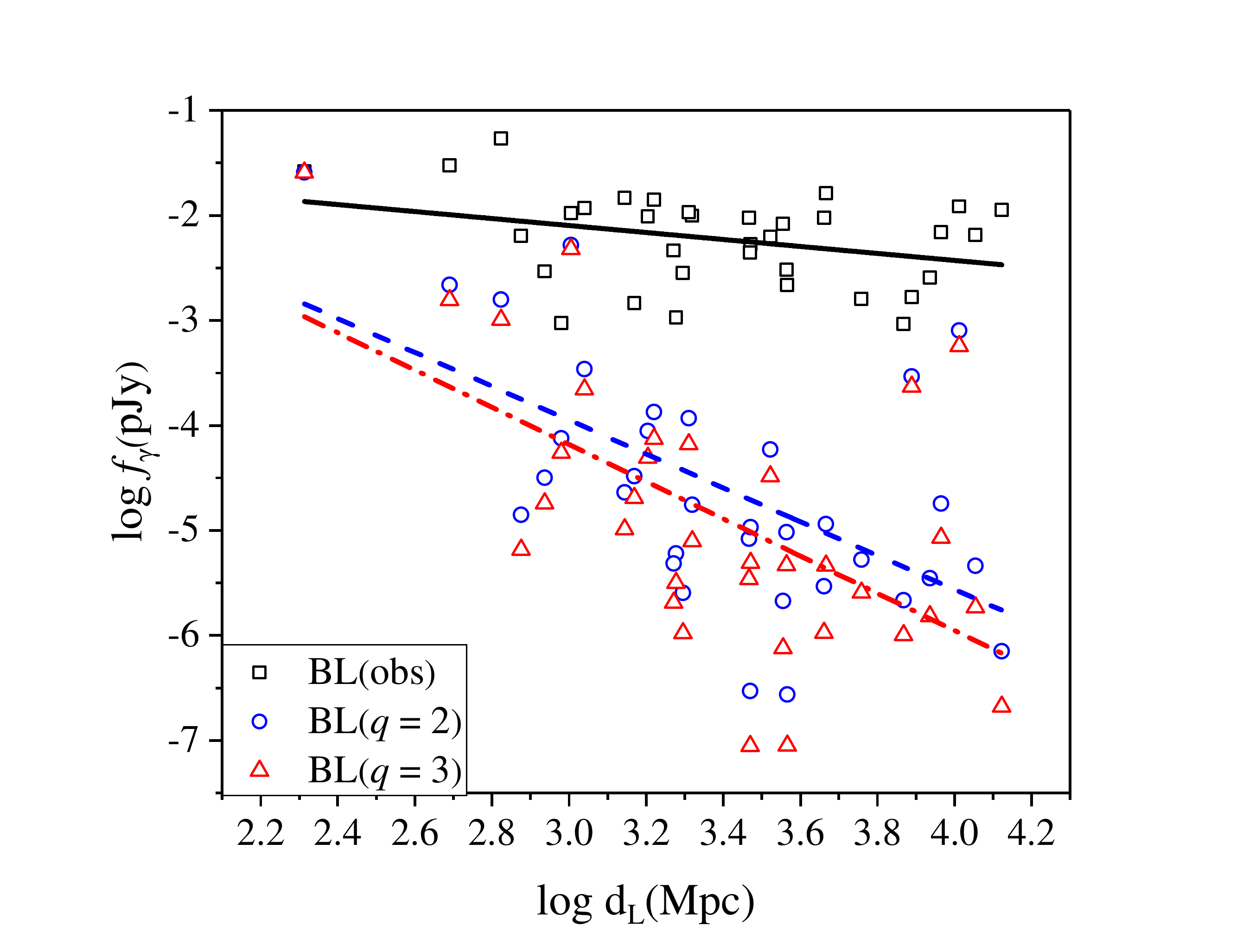}
\caption{Plot of the relations between flux densities and luminosity distances for 34 BL Lacs based on the radio Doppler factor of \citet{lio2018}. All the representations of labels are the same as Fig. \ref{fig:2}.}
\label{fig:5}
\end{figure}

As shown in  Fig. \ref{fig:1}, the distribution of intrinsic luminosity of BL Lacs is about two orders of magnitude smaller than that of the observed luminosity. If the intrinsic luminosity is assumed to be constant for BL Lacs, then one can expect a theoretical expression, $\log f_{\gamma}= -2.0\log d_{\rm{L}}  + $ const. However, the  observed $\gamma$-ray flux density does not obey to the relation, the slopes between the observed flux densities and luminosity distances for the whole sample and the subclasses are significantly different than the theoretical slope, -2. The correlation between observed $\gamma$-ray flux and redshift is also discussed in \citet{lin2011}. The nonparametric Spearman value -0.319 and probability 31\% \citep{lin2011} are suggesting that there is no correlation between observed $\gamma$-ray flux and redshift, which render the same result as our observed sample. One would naturally expect the more distant sources with fainter emission, but {\it Fermi} BL Lacs show the similar flux covering a wide range of distances. The problem is perhaps caused by the beaming effect, which results in the boosting in the observed flux density. One should use the intrinsic flux density to investigate the correlation between the flux density (de-beamed flux density) and luminosity distance.  When de-beaming effect is considered for the observed $\gamma$-ray flux density, strongly linear correlations between intrinsic emissions and luminosity distances are obtained as expected, and presented in Fig. \ref{fig:2}-\ref{fig:3} and Table \ref{tbl:2}. The slope in a continuous jet for HBLs  is closer to -2 than that in the case of a moving sphere, supporting the continuous morphology of jet emission in $\gamma$-ray band for HBLs, 
while IBLs show that the slope of $q$ = 3 is more similar to the theoretical formula, $\log f_{\gamma}= -2.0\log d_{\rm{L}}  + $ const. Limited by statistics, the results of LBLs may not be representative and valuable.

\setlength{\tabcolsep}{14pt}
\begin{table*}
\caption{Linear Correlation fitting Result. ($y=(a \pm \Delta a) \log d_{\rm{L}}+(b \pm \Delta b)$)} 
 \label{tbl:2}
\begin{tabular}{lccccccc}
\hline
\hline
$y \sim \log d_{\rm{L}}$ & $a \pm \Delta a$ & $b \pm  \Delta  b$ & $N_{\rm{A}}$& $r_{f z}$ & $p_{f z}$ & $r_{f z, \delta}$ &$p_{f z, \delta}$ \\
(1)&(2)&(3)&(4)&(5)&(6)&(7)&(8)\\
\hline
$\log f_{\rm{BL}}^{\rm{ob}}$&-0.16$\pm$0.05&-1.77$\pm$0.17&294&-0.174&0.27\%&\\
$\log f_{\rm{BL}}^{\rm{in2}}$&-1.96$\pm$0.06&2.18$\pm$0.19&294&-0.890&$<10^{-4}$&-0.538&$<10^{-4}$\\
$\log f_{\rm{BL}}^{\rm{in3}}$&-2.16$\pm$0.06&2.61$\pm$0.20&294&-0.897&$<10^{-4}$&-0.511&$<10^{-4}$\\
$\log f_{\rm{LBL}}^{\rm{ob}}$&-0.22$\pm$0.15&-1.76$\pm$0.51&27&-0.278&16.1\%&&\\
$\log f_{\rm{LBL}}^{\rm{in2}}$&-2.40$\pm$0.22&3.59$\pm$0.76&27&-0.907&$<10^{-4}$&-0.368&6.44\%\\
$\log f_{\rm{LBL}}^{\rm{in3}}$&-2.63$\pm$0.24&4.16$\pm$0.81&27&-0.912&$<10^{-4}$&-0.331&9.84\%\\
$\log f_{\rm{IBL}}^{\rm{ob}}$&-0.02$\pm$0.07&-2.24$\pm$0.24&135&-0.018&83.7\%&\\
$\log f_{\rm{IBL}}^{\rm{in2}}$&-1.67$\pm$0.08&1.46$\pm$0.26&135&-0.878&$<10^{-4}$&-0.421&$<10^{-4}$\\
$\log f_{\rm{IBL}}^{\rm{in3}}$&-1.85$\pm$0.08&1.86$\pm$0.27&135&-0.890&$<10^{-4}$&-0.397&$<10^{-4}$\\
$\log f_{\rm{HBL}}^{\rm{ob}}$&-0.26$\pm$0.08&-1.39$\pm$0.26&132&-0.274&0.15\%&\\
$\log f_{\rm{HBL}}^{\rm{in2}}$&-2.14$\pm$0.08&2.60$\pm$0.24&132&-0.928&$<10^{-4}$&-0.660&$<10^{-4}$\\
$\log f_{\rm{HBL}}^{\rm{in3}}$&-2.35$\pm$0.08&3.04$\pm$0.25&132&-0.935&$<10^{-4}$&-0.637&$<10^{-4}$\\
\hline 
\end{tabular}\label{tab2}
\medskip

\small
Notes: Col. 1 is for the linear  relation, `ob' for observed data, `in2' for a continuous jet ($q$ = 2), `in3' for a the case of a moving sphere ($q$ = 3),  Col. 2 slope ($a$) and its uncertainty ($\Delta a$); Col. 3 intercept ($b$) and its uncertainty ($\Delta b$); Col. 4 number of the sample for linear fitting; Col. 5 correlation coefficient between flux densities and luminosity distances ($r_{f d}$); Col. 6 probability ($p_{f d}$) for Col. 5;  Col. 7 correlation coefficient between flux densities and luminosity distances after removing the Doppler factor effect,   ($r_{f d, \delta}$); Col. 8 probability ($p_{f d, \delta}$) for Col. 7.

\end{table*}
The slopes for intrinsic flux density are not exactly equal to -2; these differences maybe from the strong assumption, namely that the intrinsic luminosity is the same for all BL Lacs. As shown in Fig. \ref{fig:1}, the intrinsic luminosity is mainly in a coverage of 2.5 dex, which will lead to the fitting slope deviating from the the theoretical slope of -2.
\begin{figure}[bht]
\includegraphics[width=1\columnwidth]{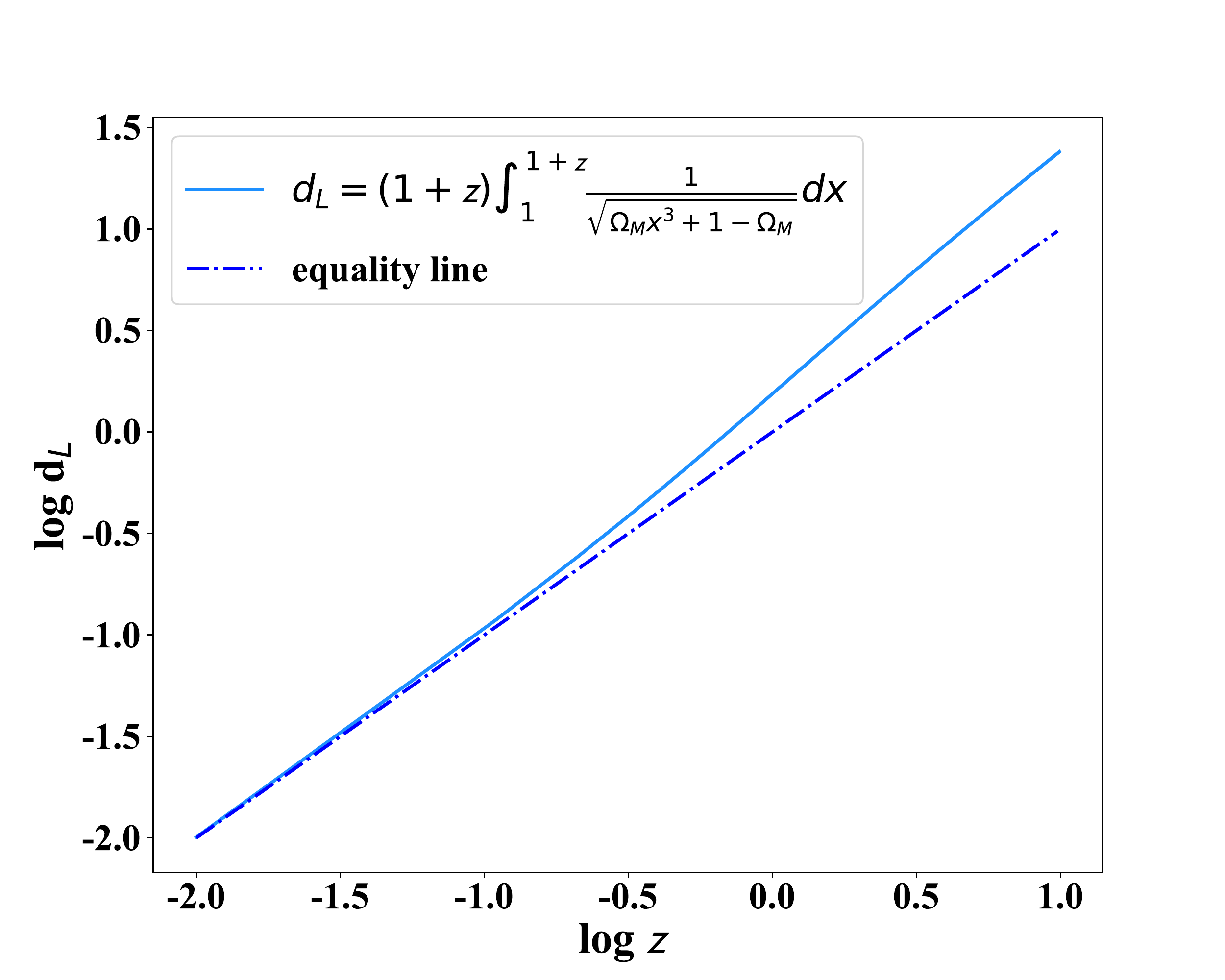}
\caption{The {\it solid line} for relation between luminosity distance and redshift. The {\it dot-dashed line} is for equality line.}
\label{fig:6}
\end{figure}

From equation (\ref{in_flux}), one can see that the intrinsic flux density is derived by the observed flux density and Doppler factor. Meanwhile, luminosity distances are close correlated with the Doppler factors (see Fig. \ref{fig:4}). It is natural to concern that the close correlations between the intrinsic flux densities and the  luminosity distances are due to the Doppler factor but not from the theoretical relationship, $\log f_{\gamma}= -2.0\log d_{\rm{L}}  + $ const. In this case, we consider the correlation by removing  the Doppler factor effect. We applied a partial correlation as did in \citet{pav1992}, 
\begin{equation*}
r_{f d, \delta}=\frac{r_{f d}-r_{f \delta} r_{d \delta}}{\sqrt{\left(1-r_{f \delta}^{2}\right)\left(1-r_{d \delta}^{2}\right)}}
\end{equation*}
the labels $f$, $d$, and $\delta$ stand for the flux density, luminosity distance, and Doppler factor respectively,  $r_{f d, \delta}$ for correlation coefficient between flux densities and luminosity distances after removing the Doppler factor effect. When the mutual correlations for the flux densities against Doppler factors, $r_{f \delta}$, and the luminosity distances against Doppler factors, $r_{d \delta}$, are computed, we can obtain the partial correlation coefficient ($r_{f d, \delta}$) and probability ($p_{f d, \delta}$), which are listed in the Col. 7-8 of Table \ref{tbl:2}.  After removing the Doppler factor effect, the correlations between intrinsic flux densities and luminosity distances still exist for the whole sample and the subclasses, except for LBLs, which give the probabilities of 6.44\% ($q$ = 2) and 9.84\% ($q$ = 3). Since the LBL sample is small, the probabilities will be easily influenced by number of sample, so their results may not be representative and will be discussed in the future with more available observed data.

Doppler factor estimation in radio band was done by \citet{lio2018}, according to the difference in the variability brightness temperature ($T_{\rm{var}}$) and the equipartition brightness temperature ($T_{\rm{eq}}$), $\delta_{\rm{var}}=(1+z)\sqrt[3]{T_{\rm{var}}/T_{\rm{eq}}}$. We collected the variability Doppler factors of \citet{lio2018} to calculate the intrinsic flux densities for {\it Fermi} BL Lacs.  The intrinsic flux densities are also closely correlated with luminosity distances for the  BL Lacs as shown in Fig. \ref{fig:5}. Although the de-beamed flux density does not show expected slope [log  $f^{\rm{in2}}_{\rm{BL}}=-(1.61\pm0.39)\log d_{\rm{L}} + (0.89\pm1.35)$ for $q$ = 2, log $f^{\rm{in3}}_{\rm{BL}}=-(1.77\pm0.44)\log d_{\rm{L}} + (1.13\pm1.49)$ for $q$ = 3]. It is possible that the larger uncertainties in Doppler factors of \citet{lio2018} and number of sample have an effect on the fitting result, the slope for instance.

This similar work was done by \citet{xiao2015} and \citet{lin2017}. \citet{xiao2015} collected 73 {\it Fermi} blazars from the {\it Fermi}/LAT second catalogue to study the correlations between $\gamma$-ray flux densities, $\log f_\gamma$, and redshift, $\log z$, and found  that the intrinsic flux densities have better correlation with redshifts and suggested the jet emissions may be the case of $q$ = 2 in $\gamma$-ray band, but the intrinsic emissions for subclass of blazars (e.g. FSRQs, BL Lacs, LBLs, HBLs) have not been discussed in \citet{xiao2015}.  \citet{lin2017} compiled a sample of 91 {\it Fermi} blazars with available radio Doppler factor, then investigated the relations between $\gamma$-ray flux densities and redshifts as shown in their Fig. 1, 2, and 3 of \citet{lin2017}. Their results supported the jet emissions  in 51 LSP blazars are continuous in $\gamma$-ray band, but the slopes for the case of a moving sphere ($q$ = 3) of 40 ISP blazars is closer to slope of -2.  However, the  blazar samples of \citet{xiao2015} and \citet{lin2017} are both small. Additionally, although the luminosity distance is obtained by redshift, when we use redshift instead of luminosity distance to discuss the correlation, it may deviate from our expected result when the redshift is large, as shown in Fig. \ref{fig:6}.  We compiled a large sample of 294 {\it Fermi} BL Lacs to explore the correlation for flux densities against luminosity distances, in which the results of our work are more reasonable than those of \citet{xiao2015} and \citet{lin2017}, who used the redshift to study  the correlation.

\section{Conclusions}\label{s:4}

BL Lacs are a sub-category of the blazar  with $\gamma$-ray emissions. From the latest Fermi catalogue 4FGL-DR2 \citep{abd2020} and BL Lacs with available Doppler factors in \citet{pei2020}, we obtained a sample of 294 BL Lacs with redshift, Doppler factor and $\gamma$-ray emissions, and investigated the correlations between the $\gamma$-ray flux densities against luminosity distances for the observed  and the intrinsic $\gamma$-ray fluxes. To get the intrinsic $\gamma$-ray flux, we considered two cases of jet, namely the case of a moving sphere and a continuous jet, and compared the result with the expected theoretical results. The conclusions are:

(1) There are close correlations between intrinsic $\gamma$-ray flux densities and luminosity distances for the whole sample and the subclasses of BL Lacs, supporting the idea that $\gamma$-ray emissions in BL Lacs are strongly beamed.

(2) From correlations between intrinsic flux densities and luminosity distances, different subclasses of blazars may have different emission morphologies. The jet emission morphology of HBLs may be continuous in $\gamma$-ray bands, while the jets emission morphology of  IBL may be a moving sphere case in the $\gamma$-ray bands.

\acknowledgments

 We thank the anonymous referee for the comments, which has helped us to improve our manuscript.


\begin{ethics}
\begin{conflict}
The authors declare no conflicts of interest.
\end{conflict}
\end{ethics}
\begin{authorcontribution}
Z. Y. Pei, J. H. Yang, J. H. Fan contributed to the study conception and design. Data collection and analysis were performed by X. T. Zeng, W. X. Yang, Y. H. Xuan, J. W. Huang, X. H. Ye and H. S. Huang. The first draft of the manuscript was written by X. H. Ye and all authors commented on previous versions of the manuscript. All authors read and approved the final manuscript.
\end{authorcontribution}

\begin{fundinginformation}
This work is supported by the National Natural Science Foundation of China under Contract No. NSFC  U2031201,  NSFC 11733001, and NSFC U2031112, and partially supported by Natural Science Foundation of Guangdong Province under Contracts No. 2019B030302001, support for Astrophysics Key Subjects of Guangdong Province and Guangzhou City, and Guangzhou University under Contract No.YM2020001.
\end{fundinginformation}
\bibliographystyle{spr-mp-nameyear-cnd}  
\bibliography{ye_refa}                
\begin{dataavailability}
The datasets of this article were derived from \citet{pei2020}.  Data will be made available on reasonable request.   Readers can also contact author to obtain the whole sample.
\end{dataavailability}

 
\end{document}